\documentclass[10pt,aps,prb,twocolumn,floatfix,showpacs,superscriptaddress]{revtex4-1}
\usepackage{graphicx,amsmath,amsfonts,mathrsfs}
\usepackage{bm}
\usepackage[]{color}

\begin{document}
\title{Tunable vertical ferroelectricity and domain walls by interlayer sliding in $\beta$-ZrI$_{2}$}
\author{Xiaonan Ma}
\affiliation{Physics Department, State Key Laboratory of Advanced Special Steel, and International Center of Quantum and Molecular Structures, Shanghai University, Shanghai 200444, China}
\affiliation{Materials Genome Institute, and Shanghai Key Laboratory of High Temperature Superconductors, Shanghai University, Shanghai 200444, China}
\author{Chang Liu}
\affiliation{Physics Department, State Key Laboratory of Advanced Special Steel, and International Center of Quantum and Molecular Structures, Shanghai University, Shanghai 200444, China}
\affiliation{Materials Genome Institute, and Shanghai Key Laboratory of High Temperature Superconductors, Shanghai University, Shanghai 200444, China}
\author{Wei Ren}
\email{renwei@shu.edu.cn}
\affiliation{Physics Department, State Key Laboratory of Advanced Special Steel, and International Center of Quantum and Molecular Structures, Shanghai University, Shanghai 200444, China}
\affiliation{Materials Genome Institute, and Shanghai Key Laboratory of High Temperature Superconductors, Shanghai University, Shanghai 200444, China}
\author{Sergey A. Nikolaev}
\email{nikolaev.s.aa@m.titech.ac.jp}
\affiliation{Tokyo Tech World Research Hub Initiative (WRHI), Institute of Innovative Research, Tokyo Institute of Technology, 4259 Nagatsuta, Midori-ku, Yokohama, Kanagawa 226-8503, Japan}
\affiliation{Laboratory for Materials and Structures, Tokyo Institute of Technology, 4259 Nagatsuta, Midori-ku, Yokohama, Kanagawa 226-8503, Japan}
\affiliation{National Institute for Materials Science, MANA, 1-1 Namiki, Tsukuba,Ibaraki 305-0044, Japan}

\date{\today}

%%%%%%%%%%%%%%%%%%%%%%%%%
\begin{abstract}
Vertical ferroelectricity where a net dipole moment appears as a result of in-plane ionic displacements has gained enormous attention following its discovery in transition metal dichalcogenides. Based on first-principles calculations, we report on the evidence of robust vertical ferroelectricity upon interlayer sliding in layered semiconducting $\beta$-ZrI$_{2}$, a sister material of polar semimetals MoTe$_{2}$ and WTe$_{2}$. The microscopic origin of ferroelectricity in ZrI$_{2}$ is attributed to asymmetric shifts of electronic charges within a trilayer, revealing a subtle interplay of rigid sliding displacements and charge redistribution down to ultrathin thicknesses. We further investigate the variety of ferroelectric domain boundaries and predict a stable charged domain wall with a quasi-two-dimensional electron gas and a high built-in electric field that can increase electron mobility and electromechanical response in multifunctional devices. Semiconducting behaviour and a small switching barrier of ZrI$_{2}$ hold promise for novel ferroelectric applications, and our results provide important insights for further development of slidetronics ferroelectricity.
\end{abstract}
\maketitle

\section{Introduction}
\par Ferroelectric materials with spontaneous electric dipole moments switchable by an external electric field offer a broad range of technological applications, such as non-volatile memories, field-effect transistors, and active elements in electromechanical and electrooptical devices.\cite{ferro1,ferro2} The ability to switch electric polarization is a key ingredient in modern nanotechnology, where the need for further reduction of individually polarized domains towards the atomic scale, as well as for the ease of their switchability has been been constantly growing. 

\par Thinning down ferroelectrics is one promising direction to miniaturize electronic devices, and several ferroelectric materials have been found to maintain macroscopic polarization in ultrathin films.\cite{thin1,thin2,thin3,thin4,thin5} Furthermore, the development of van der Waals assembly has enabled heterostructure engineering,\cite{assembly1,assembly2} where physical properties are tuned to a desired functionality by combining different individual layers, which can be used to design two-dimensional ferroelectrics from non-ferroelectric parent compounds.\cite{boron1,boron2} Other venues to overcome the challenges can be found in van der Waals layered materials that possess properties favourable for tailoring ferroelectricity, such as durability against strain and surface functionalization.\cite{vdwreview} Robust polarization in these systems was shown to sustain down to atomic thicknesses and can provide new building blocks for functional heterostructures.\cite{thin33,thin44,wte2switch,polarmote2}  

\par Following the immense growth of activity in two-dimensional systems, layered transition metal dichalcogenides (TMDs) have recently drawn great attention due to their diverse physical properties, ranging from extremely large magnetoresistance~\cite{magneto1} and superconductivity\cite{supermote2} to the topological electronic states.\cite{zhijun,higherorder} Recent studies have demonstrated that the out-of-plane switchable polarization originating from interlayer sliding exists in polar Weyl semimetals MoTe$_{2}$ and WTe$_{2}$.\cite{polarmote,wte1} Although the value of spontaneous polarization is small and can be partially screened by metallic states, its rigidity upon interlayer sliding and importance for potential applications as a ferroelectric memory have prompted the urge to search for novel vertical ferroelectrics. Despite rising research activity in this field dubbed slidetronics, only a few materials have been discovered so far, such as MoTe$_{2}$, WTe$_{2}$, CuInP$_{2}$S$_{6}$,\cite{cuinps} In$_{2}$Se$_{3}$,\cite{inse1,inse2}, and VS$_{2}$.\cite{renwei} Exploration of new layered ferroelectrics elucidating the microscopic origin of electric polarization and their intrinsic properties, such as domain walls, are in high demand for further development of slidetronics.

%%%%%%%%%%%%%%%%%%%%%%%
\begin{figure*}[t!]
\begin{center}
\includegraphics[width=0.99\textwidth]{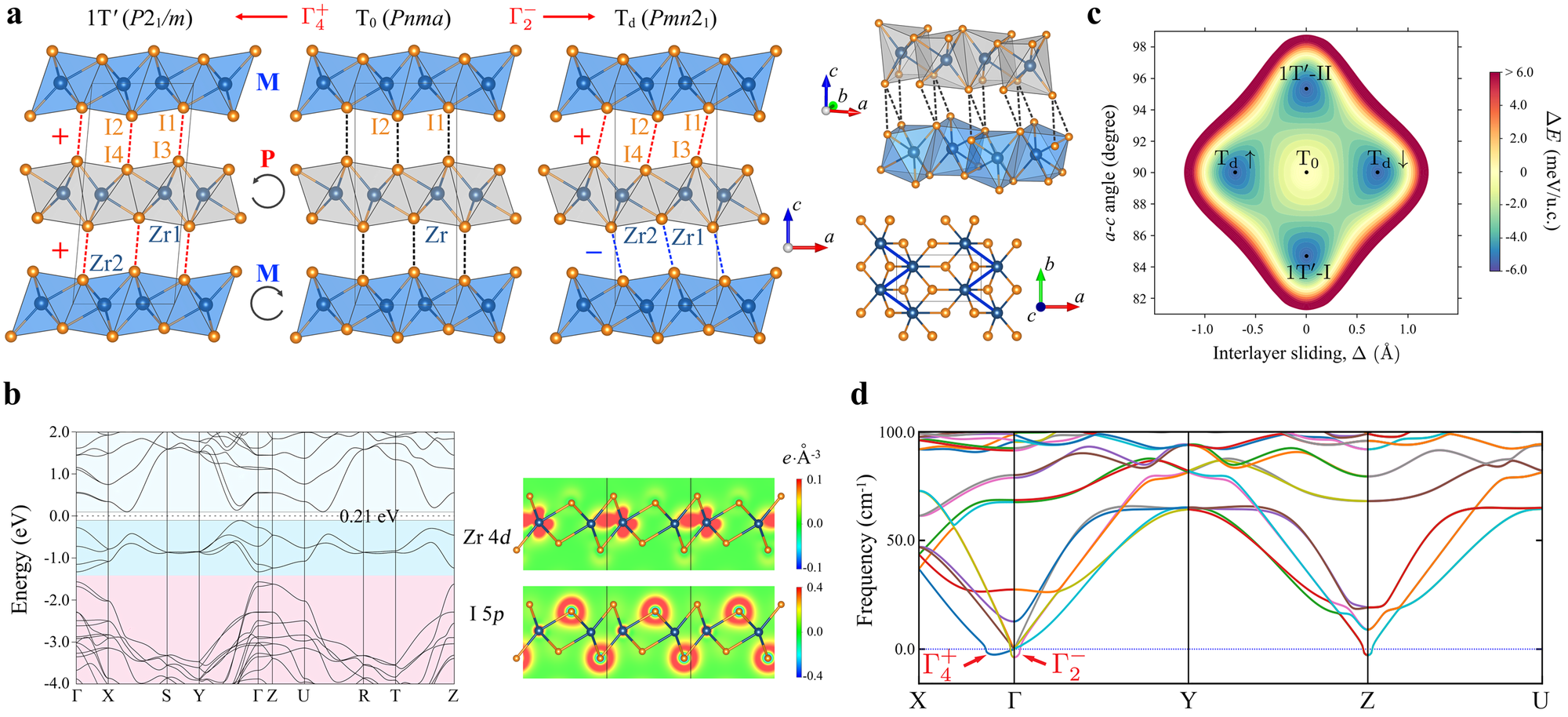}
\end{center}
\caption{\textbf{Relation of the T$_{0}$, T$_{\mathrm{d}}$ and 1T$'$ phases of ZrI$_{2}$.} \textbf{a}~Crystal structures: 1T$'$, designed T$_{0}$, and T$_{\mathrm{d}}$ phases. Side and top views presented to the right show the interlayer zigzag I--I bonds (black dotted lines) and the intralayer zigzag Zr--Zr chains (blue solid lines). Trilayers with the clockwise and counterclockwise rotated ZrI$_{6}$ octahedra are denoted as M (Minus) and P (Plus), respectively. The $+/-$ signs stand for positive and negative displacements of the interlayer zigzag I--I bonds (with respect to the lower trilayer). The $\Gamma_{2}^{-}$ and $\Gamma_{4}^{+}$ modes relate the $Pnma$ structure to the $Pmn2_{1}$ (T$_{\mathrm{d}}$) and $P2_{1}/m$ (1T$'$) phases, respectively. \textbf{b}~Band structure of the designed T$_{0}$ phase and partial charge densities. The Zr $d$ and I $p$ states are highlighted with blue and pink colours, respectively. \textbf{c}~Energy profile obtained from the devised T$_{0}$ phase as a function of interlayer displacement and the monoclinic $a-c$ angle. \textbf{d}~Enlarged phonon spectrum of the T$_{0}$ phase of ZrI$_{2}$.}
\label{fig:all}
\end{figure*}
%%%%%%%%%%%%%%%%%%%%%%%

\par In this work, we report on the first-principles evidence of robust vertical ferroelectricity in layered ZrI$_{2}$, a semiconducting counterpart of isostructural semimetal TMDs MoTe$_{2}$ and WTe$_{2}$. Being several orders of magnitude smaller than in conventional ferroelectrics, the out-of-plane polarization in ZrI$_{2}$ is found to be rigid upon interlayer sliding, and the low energy barrier for its ferroelectric switching combined with a small band gap can hold out the prospect for novel slidetronics applications. Our theoretical study reveals that the ferroelectric activity in ZrI$_{2}$ stems from a subtle interplay of charge redistribution and ionic displacements, providing important insights on the origin of vertical ferroelectricity in layered materials and justifying its persistence in the ultrathin limit. We investigate the complexity of domain boundaries in multidomain structures of ZrI$_{2}$ that arise due to the breaking of stacking sequences. Our results demonstrate the formation of stable charged domain walls with a quasi-two-dimensional electron gas and a high built-in electric field that can be put to good use in multifunctional devices.

%%%%%%%%%%%%%%%%%%%%%%%%%%%%%%%%%%
\section{RESULTS}
\subsection{Ferroelectric activity in ZrI$_{2}$}
\par Historically, there have been identified three polymorph forms of ZrI$_{2}$ studied by Guthrie and Corbett: $\alpha$ ($P2_{1}/m$)\cite{struc1}, $\beta$ ($Pmn2_{1}$)\cite{struc2}, and $\gamma$ ($R\bar{3}$)\cite{struc3} phases. The $\gamma$-phase is comprised of the antiprism clusters formed by six Zr atoms which are surrounded by twelve iodine atoms, whereas the $\alpha$- and $\beta$-phases have layered structures and are reported to be isostructural, respectively, to the 1T$'$ and T$_{d}$ phases of MoTe$_{2}$ and WTe$_{2}$. The latter phases consist of the buckled I--Zr--I trilayers coupled by weak van der Waals interactions, where the Zr atoms form the zigzag chains in each layer, as shown in Fig.~\ref{fig:all}a. Among the three polymorphs, only the $\beta$-phase has a polar structure and can be expected to reveal ferroelectric properties.
 
\par Similar to its sister compounds MoTe$_{2}$ and WTe$_{2}$, the $\alpha$- and $\beta$-phases of ZrI$_{2}$ (hereafter referred to as 1T$'$ and T$_{\mathrm{d}}$, respectively) can be derived from the devised parent T$_{0}$ phase with the $Pnma$ structure by distorting the orthorhombic $a-c$ angle or by sliding adjacent trilayers, respectively (Fig.~\ref{fig:all}a). All three phases are found to be semiconducting with an indirect energy band gap $E_{g}\sim0.20$ eV (see Supplementary Figure~1). As shown in Fig.~\ref{fig:all}b, the states near the Fermi level correspond to the Zr $d$ states that form strong metal-metal bonding along the zigzag chains, and the bands below represent the I $p$ states. In contrast to semimetals MoTe$_{2}$ and WTe$_{2}$ which demonstrate a sizeable mixing of the anionic $p$ and metal $d$ states at the Fermi level that alongside with spin-orbit coupling stabilizes the Weyl points,\cite{zhijun} the I $p$ states in ZrI$_{2}$ are pushed way below the Fermi level due to much longer intralayer I--I bonds, without affecting a small gap opened in the Zr $d$ states.

\par In Fig.~\ref{fig:all}c, we computed the energy landscape as a function of the monoclinic distortion and interlayer sliding starting from the devised T$_{0}$ structure and fixing the unit cell volume and atomic arrangement in each trilayer. For both types of displacement, a double-well structure is clearly seen in the calculated energy profile, where the center with higher energy corresponds to the parent T$_{0}$ phase. By this means, interlayer sliding leads to two local minima at $\Delta\sim\pm0.71$~\AA~representing two polar T$_{\mathrm{d}}$ structures, whereas rotating the monoclinic $a-c$ angle stabilizes two ferroelastic 1T$'$-I and 1T$'$-II phases at 84.5$^{\circ}$ and 95.5$^{\circ}$, respectively. The corresponding energy barrier connecting the twin structures via direct pathways through the T$_{0}$ phase is $\sim5.3$~meV/u.c for both the T$_{\mathrm{d}}$ and 1T$'$ phases, and the energy difference between the 1T$'$ and T$_{\mathrm{d}}$ phases is found to be small, less then 0.05 meV/u.c. ($E_{T_{\mathrm{d}}}-E_{1T'}=-0.17$ meV/u.c. for the fully optimized structures). 

\par The calculated phonon spectra of the T$_{0}$ phase presented in Fig.~\ref{fig:all}d clearly indicate two types of lattice instabilities: an optical zone-centered mode $\Gamma_{2}^{-}$ and a linear phonon mode $\Gamma_{4}^{+}$ along the $\Gamma$-X direction carrying an elastic instability. From group-theoretical analysis (see Supplementary Table~3), it follows that the two modes with irreducible representations $\Gamma_{2}^{-}$ and $\Gamma_{4}^{+}$ transform the $Pnma$ phase to the $Pmn2_{1}$ and $P2_{1}/m$ structures, respectively. Indeed, the atomic displacement corresponding to the $\Gamma_{2}^{-}$ instability include an in-plane sliding displacement of the alternating trilayers, while the $\Gamma_{4}^{+}$ mode amounts to a shear distortion of the unit cell, resulting in a ferroelastic structure.

%%%%%%%%%%%%%%%%%%%%%%%
\begin{figure}[t!]
\begin{center}
\includegraphics[width=0.49\textwidth]{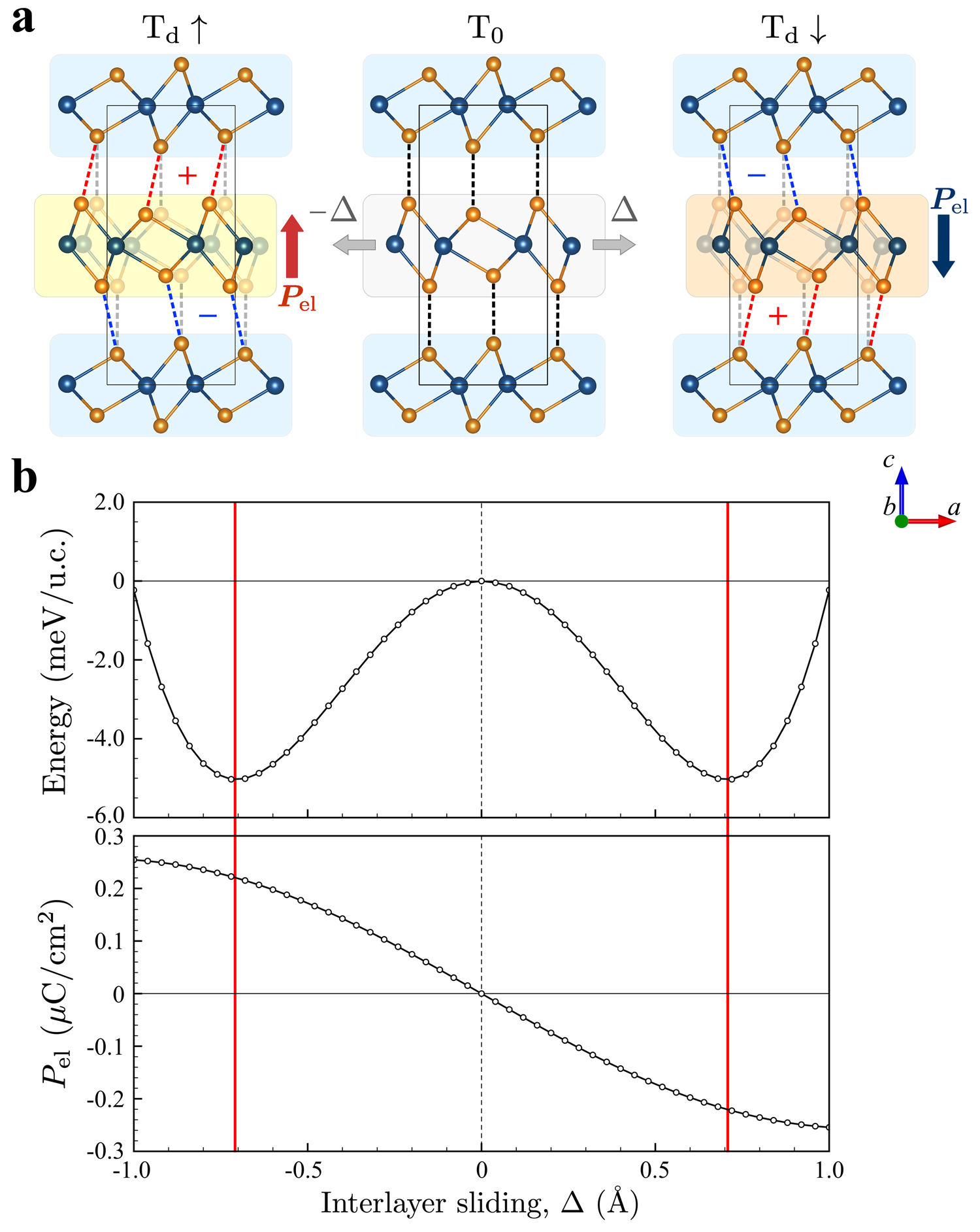}
\end{center}
\caption{\textbf{Ferroelectric polarization in the T$_{\mathrm{d}}$ phase of ZrI$_{2}$.} \textbf{a}~T$_{\mathrm{d}\uparrow}$ and T$_{\mathrm{d}\downarrow}$ structures obtained from the devised T$_{0}$ phase by interlayer sliding. The M trilayers are denoted with blue rectangles, and the P trilayers with positive and negative sliding displacements are highlighted with orange and yellow rectangles, respectively. \textbf{b}~Double-well energy profile and electric polarization along the $c$ axis calculated from the Berry phase theory as a function of interlayer sliding in the devised T$_{0}$ phase.}
\label{fig:pol}
\end{figure}
%%%%%%%%%%%%%%%%%%%%%%%

\par The T$_{0}$, 1T$'$ and T$_{\mathrm{d}}$ phases can be characterized by the way the trilayer sequences are stacked in a layered structure. In the T$_{0}$ phase with the $D_{2h}$ symmetry, the I--Zr--I trilayers are invariant under the mirror reflection with respect to the $b$ axis accompanied by a half-lattice vector translation, $\{M_{b}|\frac{\boldsymbol{b}}{2}\}$, and the adjacent trilayers can be transformed by $\{M_{a}|\frac{\boldsymbol{a}}{2}+\frac{\boldsymbol{b}}{2}\}$ or $\{M_{c}|\frac{\boldsymbol{a}}{2}\}$. Consequently, the trilayers are stacked with two alternating orientations where the ZrI$_{6}$ octahedra twist either clockwise or counterclockwise, denoted by taking notations of Ref.~\cite{polarmote} as M (Minus) and P (Plus), respectively. In the T$_{\mathrm{d}}$ phase, the interlayer sliding along the $a$ axis breaks inversion symmetry reducing the point symmetry to $C_{2v}$, so that the adjacent trilayers are invariant under $\{M_{b}|\frac{\boldsymbol{b}}{2}\}$ and can be connected by $\{M_{a}| \Delta+\frac{\boldsymbol{a}}{2}+\frac{\boldsymbol{b}}{2}\}$. On the other hand, the monoclinic distortion along the $a$ axis  in the 1T$'$ phase preserves inversion symmetry but lowers the symmetry to $C_{2h}$ with the mirror plane along the $b$ axis. The twin structures in both phases driven by the $\Gamma_{2}^{-}$ and $\Gamma_{4}^{+}$ modes can be further classified by the way the I ions in adjacent trilayers move relative to each other: in a positive ($+$) or negative ($-$) direction, if seen with respect to the lower layer, as shown in Fig.~\ref{fig:all}a. Then, if going up from the bottom trilayer, the patterns M$+$P$+$ and M$-$P$-$ will correspond to the ferroelastic 1T$'$-I and 1T$'$-II structures, while the patterns M$-$P$+$ and M$+$P$-$ will stand for the polar T$_{\mathrm{d}\uparrow}$ and T$_{\mathrm{d}\downarrow}$ phases, respectively. With that said, the 1T$'$ phase has only one type of the interlayer displacement, whereas the T$_{\mathrm{d}}$ phase contains both $+$/$-$ types that break inversion symmetry.

%%%%%%%%%%%%%%%%%%%%%%%
\begin{figure}[b]
\begin{center}
\includegraphics[width=0.49\textwidth]{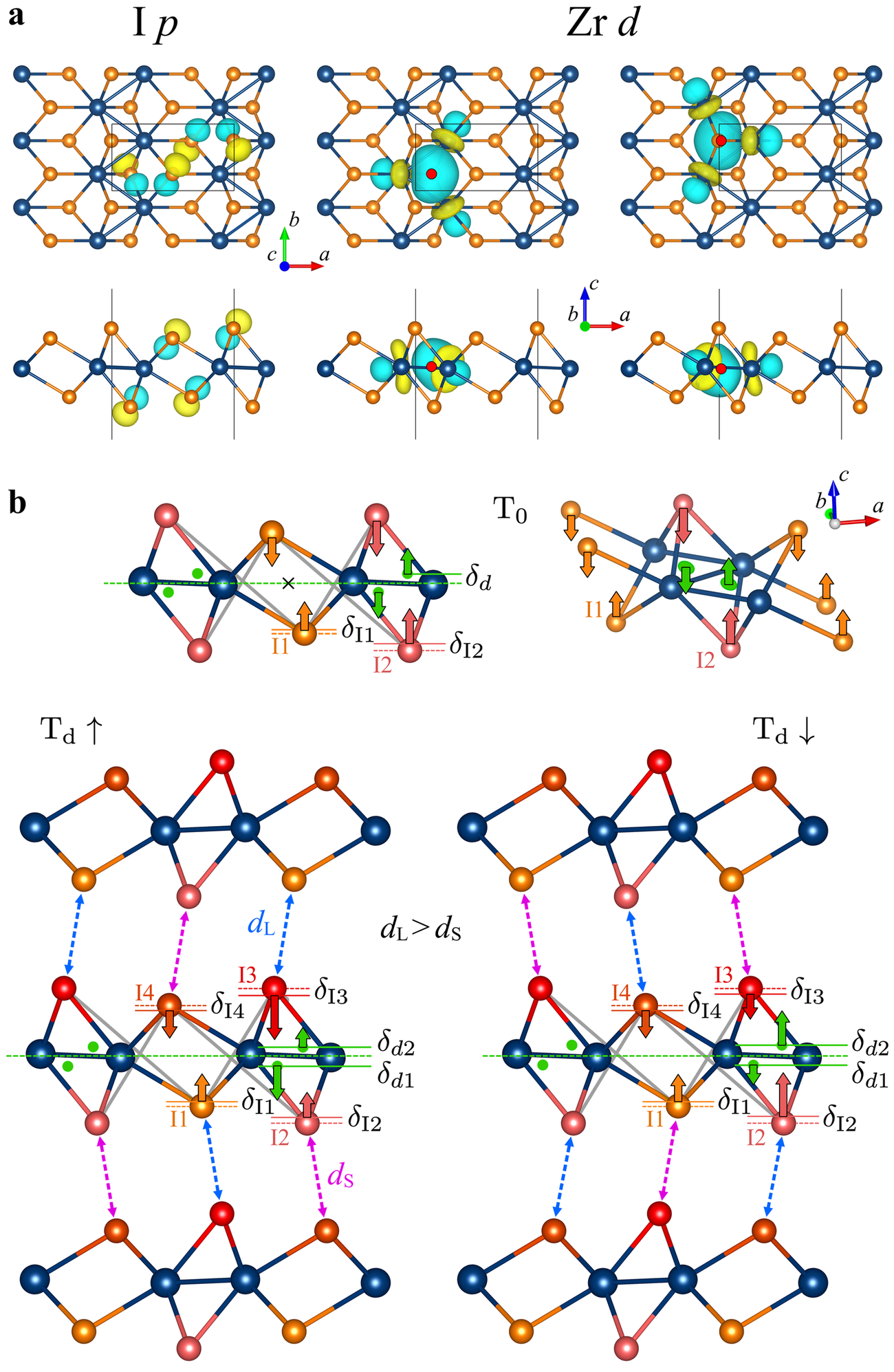}
\end{center}
\caption{\textbf{Schematics of electronic charge displacements in ZrI$_{2}$.} \textbf{a}~Wannier functions (top and side views) constructed for the occupied I $p$ (left) and Zr $d$ (middle and right) states. The Wannier functions are presented for one trilayer, and only one $p$ Wannier function at each I site is shown for clarity. Centers of the $d$ orbital Wannier functions are denoted with red circles. \textbf{b}~Electronic charge displacements in the centrosymmetric and non-centrosymmetric phases due to interlayer sliding. Vertical shifts of the Wannier centers off their specified atomic positions are schematically shown with arrows. Centers of the $d$ orbital Wannier functions are denoted with green circles. The alternating longer $d_{\mathrm{L}}$  and shorter $d_{\mathrm{S}}$ zigzag I--I bonds are shown with blue and pink colours, respectively. Inversion center is marked with the black cross.}
\label{fig:wannier}
\end{figure}
%%%%%%%%%%%%%%%%%%%%%%%

\par To investigate ferroelectric activity of the T$_{\mathrm{d}}$ phase, we considered the devised T$_{0}$ phase modulated along the $\Gamma_{2}^{-}$ mode while keeping fixed the unit cell volume and atomic arrangement in each trilayer. Electric polarization calculated from the Berry phase theory as a function of interlayer sliding is presented in Fig.~\ref{fig:pol} and shows that a net dipole moment appears strictly along the vertical axis and changes its direction within the double-well profile when going between positive and negative sliding displacements. Since interlayer sliding amounts to rigid ionic translations, electric polarization is purely of electronic origin. The absolute value of $P_{\mathrm{el}}$ corresponding to the energy minimum of the constrained T$_{\mathrm{d}}$ structure is 0.22 $\mu$C$\cdot$cm$^{-2}$ (0.24~$\mu$C$\cdot$cm$^{-2}$ for the fully optimized T$_{\mathrm{d}}$ phase). Despite being a few orders of magnitude smaller compared to conventional ferroelectrics, electric polarization in ZrI$_{2}$ is caused solely by interlayer sliding and is switchable without vertical ionic displacements. Importantly, the switching barrier between two ferroelectric T$_{\mathrm{d}}$ structures of ZrI$_{2}$ is much lower than in conventional ferroelectrics (34 meV/atom for BaTiO$_{3}$ and 67 meV/atom for PbTiO$_{3}$).\cite{rabe}

\par Given the rigidity of a single trilayer, intralayer antiferroelectric structures are found to be highly unstable relaxing to the non-polar T$_{0}$ phase and confirming the robustness of ferroelectric properties in the T$_{\mathrm{d}}$ phase. From molecular dynamics simulations, the lower bound for the Curie temperature of the T$_{\mathrm{d}}$ phase can be estimated as $\sim400$~K.

\par It is worth noting that interlayer sliding alone is not sufficient to explain the appearance of the net dipole moment in the T$_{\mathrm{d}}$ phase. Several previous studies attributed the ferroelectric activity in WTe$_{2}$ to the vertical charge transfer between adjacent trilayers that occurs upon interlayer sliding.~\cite{liju,ren1} While the Berry phase analysis presented above shows that the ferroelectric activity in ZrI$_{2}$ is purely of electronic origin, the calculated change in the charge density demonstrates that the P and M trilayers are rigid in the sense that a sheer ionic displacement is accompanied by the same intralayer shift of the electronic charges, and the corresponding change in the interlayer charge density representing interlayer bonding indicates that interlayer sliding results solely in the charge density redistribution between the trilayers (see Supplementary Figure~2). 
 
 \par The microscopic origin of the uncompensated dipole moment in the T$_{\mathrm{d}}$ phase of ZrI$_{2}$ can be elaborated by examining displacements of the electronic charges driven by interlayer sliding. The electronic contribution to the net dipole moment can be expressed as a sum of the centers of the Wannier functions, $\mathbf{r}_{n}$, as $\mathbf{d}_{el}=-2e\sum_{n}\mathbf{r}_{n}$ ($e$ is a positive electron charge). As shown in Fig.~\ref{fig:wannier}a, the Wannier functions of the I $p$ states are well localized in the vicinity of the I ions, whereas the Wannier functions corresponding to the Zr $d$ states are not centered at the Zr sites and are, instead, located at the Zr triangles, reflecting strong metallic bonding in the zigzag chains. Due to the hybridization effects, the Wannier functions move off the atomic sites, and their vertical displacements can be described by the shift of the Wannier centers relative to their symmetry specified positions, $\delta_{d}$ and $\delta_{\mathrm{I}}$, as depicted in Fig.~\ref{fig:wannier}b. As a result of the intralayer hybridization, the $p$-orbital Wannier functions at the top and bottom I sites tend to shift inwards the trilayer, which is accompanied by the shift of the $d$ orbital Wannier functions at the corresponding Zr triangles. The intralayer hybridization is strong and determines an overall displacement of electronic charge centers. However, the charge density distribution is also affected by interlayer bonding, in particular by steric I--I interactions. In the centrosymmetric phase, the top and bottom I1 and I2 sites are equivalent in each trilayer resulting in symmetric shifts of the Wannier centers, so that all corresponding $\delta$'s are equal giving zero $\mathbf{d}_{el}$. The calculated charge centers shifts for the centrosymmetric T$_{0}$ phase are $\delta_{d}=0.0712$~\AA, $\delta_{\mathrm{I1}}=0.1660$~\AA, and $\delta_{\mathrm{I2}}=0.2197$~\AA~(for $\delta_{\mathrm{I}}$ is an average over three $p$ orbitals). When sliding adjacent trilayers, inversion symmetry is broken, and the top and bottom I sites in each trilayer become non-equivalent. Notably, interlayer sliding modifies steric bonding in the interlayer zigzag I--I chains leading to the alternating shorter and longer I--I bonds that change depending on the sliding direction (Fig.~\ref{fig:wannier}b). These bonds reverse their order between adjacent trilayers, so that the top and bottom I ions in each trilayer experience asymmetric bonding environment giving rise to non-equivalent shifts of their charge centers. The shorter the bond length, the more the electronic charge density is pulled outwards the trilayer reducing its vertical shift. Since interlayer bonding is weak, the resulting shifts and the net dipole moment are small. The calculated shifts in the T$_{\mathrm{d}\uparrow}$ phase are $\delta_{d1}=-0.0756$~\AA, $\delta_{d2}=0.0687$~\AA, $\delta_{\mathrm{I1}}=0.1661$~\AA, $\delta_{\mathrm{I2}}=0.2176$~\AA, $\delta_{\mathrm{I3}}=-0.2201$~\AA, $\delta_{\mathrm{I4}}=-0.1662$~\AA, and the net dipole moment is $0.0585$~$e\cdot$\AA, in perfect agreement with the value of $0.0579$~$e\cdot$\AA~($0.243$ $\mu$C$\cdot$cm$^{-2}$) obtained from the Berry phase theory. It should be noted that in the 1T$'$ phase, the interlayer zigzag I--I chains also alternate between the trilayers, but the corresponding top and bottom I ions are always connected by the same type of the I--I bonds and have symmetric shifts of their charge centers (see Supplementary Figure~3).
 
\par The hybridization also causes the in-plane shifts of the charge centers within a trilayer. Since the interlayer zigzag I--I chains alternate periodically in the bulk structure, the in-plane shifts will sum up to zero. Nevertheless, in the case of only two trilayers, such cancellation will not be perfect giving rise to the in-plane electric polarization (see Supplementary Figure~9).\cite{prbzri2}

\subsection{Domain walls}
\par When cooled down below the transition temperature, ferroelectric materials tend to form complex multidomain structures with different orientations of macroscopic polarization, and nanometric-scale domain walls (DW) develop at the domain boundaries. Structural and functional properties of DWs, such as width and energy barrier across the DW, substantially impact polarization switching process and the DW mobility.\cite{dwreview}

%%%%%%%%%%%%%%%%%%%%%%%
\begin{figure*}[t]
\begin{center}
\includegraphics[width=0.99\textwidth]{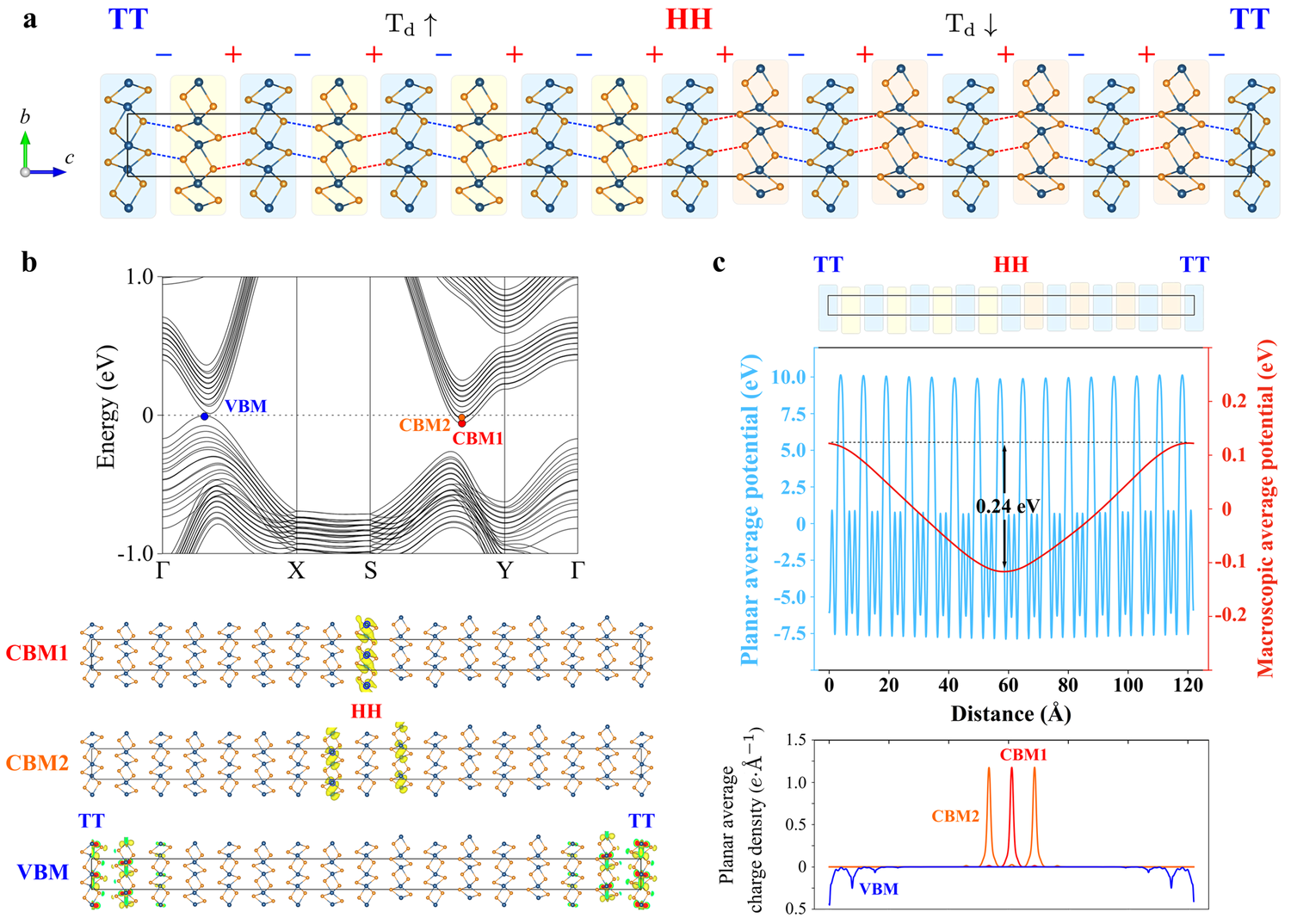}
\end{center}
\caption{\textbf{First-principles calculations of the charged DW in $\beta$-ZrI$_{2}$.} \textbf{a}~a $1\times1\times8$ supercell with the T$_{\mathrm{d}}$/T$_{\mathrm{d}}$ arrangement containing the head-to-head (HH) and tail-to-tail (TT) DWs at the center and boundaries, respectively. The M trilayers are denoted with blue rectangles, and the P trilayers with positive and negative sliding displacements are highlighted with orange and yellow rectangles, respectively. \textbf{b}~Electronic band structure and charge densities corresponding to the valence band maximum (VBM) and conduction band minimum (CBM1 and CBM2). The isosurfaces correspond to 0.01 $e\cdot$\AA$^{-3}$. \textbf{c}~Planar and macroscopic average electrostatic potentials across the supercell and planar average charge densities of VBM and CBM.}
\label{fig:dw}
\end{figure*}
%%%%%%%%%%%%%%%%%%%%%%%  

\par As follows from the results of first-principles calculations, the energy difference between the T$_{\mathrm{d}}$ and 1T$'$ phases of ZrI$_{2}$ is small, and both forms can coexist in a sample producing different types of DW structures. Following the symmetry analysis reported in Ref.~\cite{polarmote}, a stacking sequence $($T$_{\mathrm{d}})_{n}/($1T$')_{m}$ has the $Pm$ space symmetry ($C_{1h}$ point group) that can be derived from both the $P2_{1}/m$ and $Pmn2_{1}$ space groups by taking the corresponding $\Gamma_{2}^{-}$ and $\Gamma_{3}$ modes, respectively, which represent the breaking of the $+/-$ patterns. Owing to the rigidity of a single trilayer, there exist several possible ways of vertical stacking at the boundaries, which adopt the patterns of the T$_{\mathrm{d}}$ and 1T$'$ stacking sequences. For example, when two ferroelectric T$_{\mathrm{d}\uparrow}$ and T$_{\mathrm{d}\downarrow}$ domains are stacked along the $c$ axis ($...$M$-$P$+$/M$+$P$-$$...$), the 1T$'$-I pattern will be formed at the boundary. Or, similarly, for a 1T$'$-I/1T$'$-II stacking ($...$M$+$P$+$/M$-$P$-$$...$) the DW will have the T$_{\mathrm{d}\uparrow}$ pattern. Because the $Pm$ space group is polar, there are two possibilities for vertical DWs in ZrI$_{2}$: a (polar) DW with the T$_{\mathrm{d}}$ pattern at the 1T$'$/1T$'$ stacking will exhibit ferroelectric properties due to the local inversion symmetry breaking at the boundary, whereas the T$_{\mathrm{d}}$/T$_{\mathrm{d}}$ and T$_{\mathrm{d}}$/1T$'$ arrangements will form a (non-polar) charged DW with the 1T$'$ pattern arising from polarization discontinuity. 

\par When a normal component of electric polarization changes across the domain boundary as in the latter case of the T$_{\mathrm{d}}$/T$_{\mathrm{d}}$ and T$_{\mathrm{d}}$/1T$'$ arrangements, the interface will accommodate a high density of bound charges that give rise to large depolarizing fields suppressing ferroelectricity. If these charges are compensated by free carriers or electron-hole transfer across the gap, a stable charged DW can  form as a movable ultrathin conductive layer. Taking a nominal value of $P_{\mathrm{el}}$ in the T$_{\mathrm{d}}$ phase of ZrI$_{2}$ and assuming a DW width $w$ of several nanometers, the bound charge concentration can be estimated as $2P_{\mathrm{el}}/ew\sim3\cdot10^{19}$ cm$^{-3}$, which is an appreciable value for a small band gap semiconductor.\cite{cdwreview} 

\par The results obtained for the T$_{\mathrm{d}}$/T$_{\mathrm{d}}$ arrangement are summarized in Fig.~\ref{fig:dw}, where the $\mathrm{P}+\mathrm{M}+$ and $\mathrm{P}-\mathrm{M}-$ patterns at the boundaries form, respectively, the 180$^{\circ}$ head-to-head (T$_{\mathrm{d}\uparrow}$/T$_{\mathrm{d}\downarrow}$) and tail-to-tail (T$_{\mathrm{d}\downarrow}$/T$_{\mathrm{d}\uparrow}$) charged DWs. Due the presence of bound charges at the interfaces, the bottom of the conduction states and the top of the valence states approach the Fermi level, thus providing electrons and holes for screening at the head-to-head and tail-to-tail DWs, respectively (Fig.~\ref{fig:dw}b). This band bending driven by electron-hole transfer across the gap further leads to a potential difference. The calculated potential profile across the heterostructure reveals a small barrier of $\sim0.24$~eV with the minimum and maximum corresponding to the head-to-head and tail-to-tail DWs (Fig.~\ref{fig:dw}c). Assuming that a domain width can be varied up to tens of nanometers, the induced potential will give rise to a large built-in electric field of hundreds kV$\cdot$cm$^{-1}$ across the domain. Combined with a strong dielectric anisotropy of ZrI$_{2}$ (see Supplementary Table~4), the dielectric and piezoelectric responses in ZrI$_{2}$ can be greatly enhanced and controlled by changing the density of charged DWs.\cite{enhanced} The calculated formation energy of the charged DWs well converges to the estimation $E_{\mathrm{DW}}=2P_{\mathrm{el}}E_{g}/e\sim1.0$~mJ$\cdot$m$^{-2}$ as a function of the supercell size (see Supplementary Figure~11). This value is significantly smaller than in conventional ferroelectrics (35~mJ$\cdot$m$^{-2}$ for the 90$^{\circ}$ DWs and 132~mJ$\cdot$m$^{-2}$ for the 180$^{\circ}$ DWs in PbTiO$_{3}$,\cite{dw1} 71~mJ$\cdot$m$^{-2}$ for the 180$^{\circ}$ DW in BiFeO$_{3}$\cite{dw2}, $\sim70$~mJ$\cdot$m$^{-2}$for anti-phase DWs in SmFeO$_{3}$\cite{dw3}), allowing for the ease to control the DW motion (see Supplementary Figure~12).

\par As a direct consequence of strong charge compensation, the charged DWs exhibit metallic behaviour and can feature noticeable inwall conductivity.\cite{quasigas} The tail-to-tail DW at the T$_{\mathrm{d}\downarrow}$/T$_{\mathrm{d}\uparrow}$ interface is found to be less pronounced slowly decaying over several layers beyond its core, because the band bending by the valence states is rather weak (Fig.~\ref{fig:dw}c). On the contrary, the head-to-head DW at the T$_{\mathrm{d}\uparrow}$/T$_{\mathrm{d}\downarrow}$ stacking demonstrates a much higher concentration of bound charges that are strongly localised at the potential cavity of the interfacial trilayers forming a quasi-two-dimensional electron gas. Since the bound charges at the interface are proportional to the normal component of electric polarization, varying conductivities are expected for different types of DWs at the T$_{\mathrm{d}}$/T$_{\mathrm{d}}$ and T$_{\mathrm{d}}$/1T$'$ boundaries. Thus, the electron confinement can be modulated by interlayer sliding that allows to controlling both the conductivity and carrier mobility at the interfaces.\cite{gas1}

\par Taking the estimated charge density at the DW boundary, one can calculate the Debye length to be $\sim0.5$~nm at room temperature. The value is rather small, but this estimate should be taken with care. As shown in Fig.~\ref{fig:dw}, the charged DWs of ZrI$_{2}$ reveal a generic quantum structure where the electron motion perpendicular to the DW is quantized. Since the number of subbands corresponding to the electron's inwall propagation is small, the semiclassical approximations for screening effects can be violated. This feature is different from DWs in typical oxide ferroelectrics, where the number of subbands is much larger and the Thomas-Fermi approximation holds true.\cite{debyelength} 

%%%%%%%%%%%%%%%%%%%%%%%
\begin{figure}[t!]
\begin{center}
\includegraphics[width=0.49\textwidth]{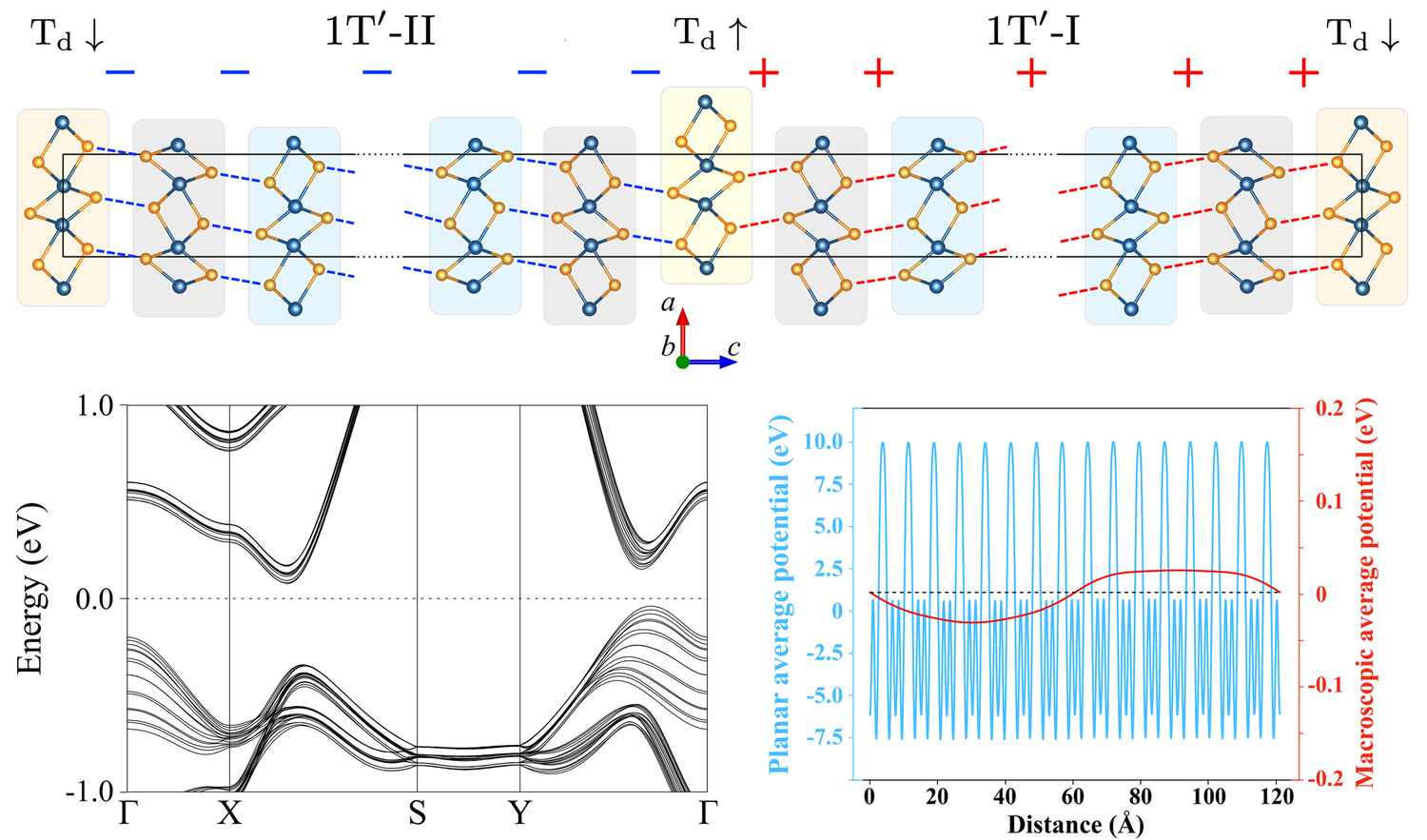}
\end{center}
\caption{\textbf{Polar DW in ZrI$_{2}$.} Electronic band structure and planar average electrostatic potential for a $1\times1\times8$ supercell with the 1T$'$-II/1T$'$-I arrangement. The M and P trilayers are denoted with blue and grey rectangles. The yellow and orange rectangles denote the T$_{\mathrm{d}\uparrow}$ and T$_{\mathrm{d}\downarrow}$ boundaries, respectively.}
\label{fig:dw2}
\end{figure}
%%%%%%%%%%%%%%%%%%%%%%%

\par An example of the polar DW structure is presented in Fig.~\ref{fig:dw2} for the 1T$'$/1T$'$ arrangement. The T$_{\mathrm{d}}$ pattern formed at the boundary breaks inversion symmetry, and the $+/-$ stacking discontinuity gives rise to the uncompensated dipole moment at the interfacial trilayer. Since the 1T$'$ phase in the bulk is non-polar, there are no boundary charges, and the 1T$'$/1T$'$ stacking remains semiconducting with the ferroelectric interface. 

\par We have considered the variety of DW structures in ZrI$_{2}$ that can be promising for novel slidetronics applications. The presented analysis is limited to the vertical arrangements and based on the breaking of stacking sequences. However, it is worth noting that the DW structures have also been reported for the in-plane boundaries in MoTe$_{2}$.\cite{polarmote} While our calculations show that the antiferroelectric arrangements in a single trilayer are highly unstable (see Supplementary Figure~14), the formation of neutral DWs at the T$_{\mathrm{d}}$/T$_{\mathrm{d}}$ structures in ZrI$_{2}$ is also possible by introducing a minimal strain mismatch between adjacent domains.

\section{Discussions}
\par Based on extensive first-principles calculations, we have established that vertical ferroelectricity in $\beta$-ZrI$_{2}$ is a robust consequence of lattice instability resolved by interlayer sliding between the adjacent trilayers. The microscopic origin of electric polarization in ZrI$_{2}$ is attributed to a subtle interplay of ionic displacements and charge redistribution, leading to asymmetric shifts of the electronic charge centers within each trilayer and thus refuting the previously proposed scenario of interlayer charge transfer. While the out-of-plane polarization in ZrI$_{2}$ is found to be several orders of magnitude smaller than in conventional ferroelectrics, the low energy barrier for its ferroelectric switching combined with a small band gap can provide opportunities for further developments in the field of ferroelectric semiconductors, such as ferroelectric tunnel junctions.\cite{junction} 

\par Our study can shed light on some aspects of ferroelectric activity in isostructural TMDs MoTe$_{2}$ and WTe$_{2}$, whose properties are summarized in Table~I.\cite{expmote2,expwte2,expwte22} The essential difference between these polar semimetals and semiconducting ZrI$_{2}$ is that electric polarization of the latter is not screened by metallic states, and $\beta$-ZrI$_{2}$ itself can be regarded as a new alternative to the layered TMDs, thus enriching the family of slidetronics ferroelectrics. The presented analysis of ferroelectric activity applies to semimetallic TMDs, explaining the persistence of electric polarization down to the ultrathin limit, and, in principle, can be extended to other vertical ferroelectrics.

\par We have established the complexity and variability of domain boundaries in multidomain structures of ZrI$_{2}$. Importantly, the same types of DWs and superlattice-like arrangements were previously reported experimentally in MoTe$_{2}$\cite{polarmote} between the Weyl semimetal T$_{\mathrm{d}}$ and higher-order topological 1T$'$ phases,\cite{zhijun,higherorder} which can provide a proving ground for exploring interfacial topological states and related quantum phenomena. Given their semimetallic behaviour, no static charge accumulation is expected at the boundaries in TMDs. In contrast, the DWs in semiconducting ZrI$_{2}$ offer another diversity related to the breaking of stacking sequences and polarization discontinuity, that greatly adds up to multifunctional aspects of vertical ferroelectrics. In particular, we have predicted a stable charged DW at the phase boundaries with a pronounced metallic behavior and a high built-in electric field that can be used to enhance dielectric and piezoelectric properties. A head-to-head charged DW was shown to form a high density quasi-two-dimensional electron gas confined at the interfacial trilayers. Such DWs can be manipulated in a controllable way and allow to creating ultrathin conductive layers embedded in a semiconducting matter and to increase electron mobility in electronic devices, that can bring new functionalities for future slidetronics applications.

%%%%%%%%%%%%%%%%%%%%%%%%%%%%
\begin{table}[b]
\caption{Experimental lattice parameters and theoretically predicted electric polarization for the T$_{d}$ phase of ZrI$_{2}$, MoTe$_{2}$, and WTe$_{2}$. }
\begin{center}
\begin{tabular}{c|ccc}
\hline
\hline
& ZrI$_{2}$ & MoTe$_{2}$ & WTe$_{2}$ \\
\hline
$a$ (\AA) & 3.744\cite{struc2} & 3.468\cite{expmote2}, 3.458\cite{zhijun} & 3.496\cite{expwte2}, 3.477\cite{expwte22} \\
$b$ (\AA) & 6.831\cite{struc2} & 6.310\cite{expmote2}, 6.304\cite{zhijun} & 6.282\cite{expwte2}, 6.249\cite{expwte22} \\
$c$ (\AA) &14.886\cite{struc2} & 13.861\cite{expmote2}, 13.859\cite{zhijun} & 14.070\cite{expwte2}, 14.018\cite{expwte22} \\
$P_{\mathrm{el}}$ ($\mu$C$\cdot$cm$^{-2}$) & 0.243 & 0.058\cite{polarmote2} & 0.190\cite{wte1} \\
Band gap (eV) & 0.20 &-- & -- \\
\hline
\hline
\end{tabular}
\end{center}
\label{tab:param}
\end{table}
%%%%%%%%%%%%%%%%%%%%%%%%%%%%

\section{Methods}
\par Structural parameters of the $Pmn2_{1}$ and $P2_{1}/m$ phases of ZrI$_{2}$ were adopted from previous experimental studies,\cite{struc1,struc2} and the $Pnma$ phase was devised by symmetrising the $Pmn2_{1}$ structure. One should bear in mind that notations of the $Pnma$ and $Pmn2_{1}$ structures have different orders of the lattice vectors, which are related as $(a,b,c)$ and $(b,a,c)$, respectively.
\par Group-theoretical analysis was performed using the Bilbao Crystallographic Server (see Supplementary Table~3).\cite{bilbao1, bilbao2} The crystal structures were visualized with \texttt{VESTA}.\cite{vesta}

\subsection{Electronic structure}
\par First-principles calculations were performed using the Vienna ab-initio simulation package (\texttt{VASP})~\cite{vasp} within the framework of projected augmented waves \cite{paw} and Quantum-ESPRESSO (\texttt{QE}) realized in the basis of plane waves.\cite{qe} The calculations were carried out using local density approximation (LDA)~\cite{lda} and generalized gradient approximation (Perdew-Burke-Ernzerhof, PBE,\cite{pbe} and Perdew-Burke-Ernzerhof revised for solids, PBEsol\cite{pbesol}) for the exchange-correlation potential. The calculations in \texttt{QE} were performed with ultrasoft and norm-conserving pseudopotentials.\cite{corso} For all calculations, the valence state configurations were taken as $4s^{2}4p^{6}5s^{2}4d^{2}$ for Zr and $5s^{2}5p^{5}$ for I. The plane wave cutoff was set to 500 eV and 900 eV for \texttt{VASP} and \texttt{QE}, respectively. The Brillouin zone was sampled by a Monkhorst-Pack $k$-point mesh,\cite{mp} $6\times10\times3$  for the T$_{0}$ and 1T$'$ phases and $10\times6\times3$ for the T$_{\mathrm{d}}$ phase. The convergence criteria for the total energy calculations was set to 10$^{-8}$ eV, and all structures were optimized with the total force convergence criteria of 10$^{-8}$~eV$\cdot$\AA$^{-1}$. The optimized crystal structures used in the main text are given in Supplementary Table~1. Electronic band structures of the T$_{0}$, T$_{\mathrm{d}}$, and 1T$'$ phases calculated with LDA, PBE and PBEsol are shown in Supplementary Figure~1. The effect of spin-orbit coupling was shown to give minor changes (see Supplementary Figure~4a). 
\par To investigate the effect of van der Waals interactions on the structural and ferroelectric properties of ZrI$_{2}$, we performed a comparative analysis using the DFT-D2 correction method of Grimme,\cite{grimme1} the DFT-D3 method with Becke-Jonson damping,\cite{grimme2,grimme3} the vdW-DF and vdW-DF2 functionals,\cite{dion1,dion2,dion3}, and the optimized vdW functionals (optPBE, optB88, and optB86b).\cite{optdion1,optdion2} The optimized structure parameters for the T$_{\mathrm{d}}$ phase are shown in Supplementary Table~2, and the corresponding band structures are presented in Supplementary Figure~5. One can see that most of the functionals show a reasonable agreement with the experimental crystal structure parameters\cite{struc2} and with the band gap of $\sim0.1$~eV\cite{struc1} (available for the 1T$'$ phase only). Significant deviations were obtained for the PBE and vdW-DF functionals that give larger values of the $c$ axis and band gaps with smaller values of electric polarization compared to the other functionals. 
\par The effect of electronic correlations in the Zr $d$ shell was checked within the LDA$+U$ method\cite{ldau} and shown to give minor changes (see Supplementary Figure~4b). We have also considered the effect of long-range Coulomb interactions using the $GW$ method\cite{gw1} as implemented in \texttt{VASP}:\cite{gw2} a single shot $G0W0$ approach with 150-200 unoccupied bands and 100-150 frequencies. The results are presented in Supplementary Figure~5. While the $GW$ method may seem applicable owing to the large spatial extension of the Zr $4d$ orbitals, one can see that the band gap is largely overestimated compared to the experimental estimates, which could be related to the implementation or other known issues of the $GW$ method in polar materials.\cite{gw3}
\par \emph{Given its computational flexibility, PBEsol without spin-orbit coupling was used for all the results presented in the main text, including the analysis of ferroelectric properties and DW structures.}
\par High-symmetry $k$-points used for the band structure and phonon dispersion calculations are $\mathrm{X}=(\frac{1}{2},0,0)$, $\mathrm{S}=(\frac{1}{2},\frac{1}{2},0)$, $\mathrm{Y}=(0,\frac{1}{2},0)$, $\mathrm{Z}=(0,0,\frac{1}{2})$, $\mathrm{U}=(\frac{1}{2},0,\frac{1}{2})$, $\mathrm{R}=(\frac{1}{2},\frac{1}{2},\frac{1}{2})$, $\mathrm{T}=(0,\frac{1}{2},\frac{1}{2})$ for the $Pnma$ (\#62) and $Pmn2_{1}$ (\#31) structures; $\mathrm{Z}=(0,\frac{1}{2},0)$, $\mathrm{B}=(0,0,\frac{1}{2})$, $\mathrm{Y}=(\frac{1}{2},0,0)$, $\mathrm{C}=(\frac{1}{2},\frac{1}{2},0)$, $\mathrm{D}=(0,\frac{1}{2},\frac{1}{2})$, $\mathrm{A}=(-\frac{1}{2},0,\frac{1}{2})$, $\mathrm{E}=(-\frac{1}{2},\frac{1}{2},\frac{1}{2})$ for the $P2_{1}/m$ (\#11) structure.

\subsection{Phonon spectra}
\par The phonon spectra for the T$_{0}$ phase of ZrI$_{2}$ were calculated using the method of frozen phonons as implemented in \texttt{VASP} and Phonopy~\cite{phonopy} and density functional perturbation theory\cite{dfpt} (DFPT) as implemented in \texttt{QE}. The calculations within frozen phonons were carried out for a $2\times4\times1$ supercell on a $3\times3\times3$ $k$-point mesh. The calculations within DFPT were performed on a $3\times6\times2$ $k$-point mesh. The calculated phonon spectra are summarized in Supplementary Figure~6.
\par The Born effective charges and the dielectric matrix with and without local field effects were calculated using DFPT as implemented in \texttt{VASP}.\cite{bornvasp}

\subsection{Electric polarization}
\par Electric polarization was calculated within the Berry phase theory.~\cite{berry1,berry2} For the results shown in Fig.~\ref{fig:pol}, we have considered the T$_{0}$ phase modulated along the $\Gamma_{2}^{-}$ mode while keeping fixed the unit cell volume and atomic arrangement in each trilayer. Electric polarization of the fully optimized T$_{\mathrm{d}}$ phase obtained within different approximations for the exchange-correlation potential is summarized in Supplementary Table~2 and Supplementary Figure~7. \par The analysis of electronic charge centers was carried out using maximally localized Wannier functions as implemented in the Wannier90 package.\cite{wan90} The resulting Wannier functions are obtained by projecting the states below the Fermi level onto the Zr $d$ and I $p$ atomic orbitals. Contributions from all sites to the net dipole moment can also be seen in the non-diagonal components of the Born effective charge tensors (see Supplementary Figure~8).

\subsection{Molecular dynamics}
\par  Ab-initio molecular dynamics simulations were performed on a $2\times2\times1$ supercell for the structure optimized with PBEsol. The NVT ensemble with the Nos\'e-Hoover thermostat\cite{nose} was chosen to simulate the effects of temperature, and the calculations were run up to 2000 fs with a time step of 2 fs. 
\par In order to estimate the Curie temperature of ZrI$_{2}$ while taking into account the energetic proximity of the T$_{\mathrm{d}}$ and 1T$'$ phases, one has to consider large supercells and allow for elastic changes in the unit cell's shape and volume. In this study, molecular dynamics was performed with the fixed volume of the supercell, and a possible structural instability of the T$_{\mathrm{d}}$ phase was associated with the deviation of the interlayer I-I distances from their values at 0~K. The results are summarized in Supplementary Figure~10. At low temperatures, the interlayer I-I bonds between the middle and upper/lower trilayers evolve on average around their equilibrium 0~K values. At temperatures higher than 400~K, the middle trilayer starts to float alternately between the upper and lower trilayers (the values of the interlayer I-I bonds with the upper and lower trilayers decrease and increase by turns). While this behaviour does not unambiguously imply the transition to the paraelectric T$_{0}$ phase, it can be associated with the onset of structural instability. We believe that such approach allows us to consider 400~K as an effective lower bound for the Curie temperature in the T$_{\mathrm{d}}$ phase of ZrI$_{2}$.

\subsection{Domain walls}
\par Calculations for the T$_{\mathrm{d}\uparrow}$/T$_{\mathrm{d}\downarrow}$ heterostructure were performed on a $1\times1\times8$, $1\times1\times10$, and $1\times1\times12$ supercells in the $Pmn2_{1}$ notation. Calculations for the 1T$'$/1T$'$ heterostructure were carried out for a $1\times1\times8$ supercell in the $Pnma$ notation. The Brillouin zone was sampled by a $10\times5\times1$ Monkhorst-Pack $k$-point mesh, and the heterostructures were optimized with the total force convergence criteria of 10$^{-5}$~eV/\AA. The DW energy is calculated as $E_{\mathrm{DW}}=(E_{\mathrm{MD}}-E_{\mathrm{SD}})/2A$, where $E_{\mathrm{MD}}$ is the energy of a supercell containing the DW, $E_{\mathrm{SD}}$ is the corresponding single domain energy, and $A$ is the DW cross-sectional area. To reduce systematic errors, the single domain structures were optimized for the same supercell. Since the $($T$_{\mathrm{d}})_{n}($1T$')_{m}$ heterostructure has the $Pm$ symmetry, there may be a tendency towards monoclinic distortions during structural optimization. We have compared our results for the fully optimized structures and the constrained structures with a fixed shape and found only minor changes in the calculated DW energies and electrostatic potentials. The results obtained with PBEsol are summarized in Supplementary Figure~11, showing the convergence of the DW energy and electrostatic potentials with the respect to the supercell size.

\par  The formation of charged DWs largely depends on the values of the band gap and electric polarization, which specify the degree of the band bending. Electronic spectra of the T$_{\mathrm{d}\uparrow}$/T$_{\mathrm{d}\downarrow}$ heterostructure obtained for a variety of exchange-correlation functionals considered in our study are compared in Supplementary Figure~13. One can see that the band bending by the conduction states and the formation of the head-to-head DWs are well defined and pronounced for most of the functionals. In contrast, the band bending from the valence states is rather weak, and the formation of the tail- to-tail DWs is less definite and may be unfavourable. The optPBE and vdW-D2 functionals only reveal the head-to-head DW, which can be attributed to the reduced value of electric polarization compared to the other functionals. The calculations with PBE and vdW-DF do not show the formation of any charged DWs, which can be related to the fact that these functionals largely overestimate the $c$ axis and give larger band gaps with smaller values of electric polarization compared to the other functionals. 
\par The Debye length was estimated using the Debye-H\"uckel model:\cite{debye}
\begin{equation}
\lambda=\sqrt{\frac{\epsilon_{r}\epsilon_{0}k_{B}T}{n e^{2}}},
\end{equation}
\noindent where $\epsilon_{r}$ is the static dielectric constant along the $c$ axis, $\epsilon_{0}$ is the vacuum permittivity, $k_{B}$ is the Boltzmann constant, $T$ is temperature, $n$ is the carrier charge density at the DW, and $e$ is the elementary charge.

%%%%%%%%%%%%%%%%%%%%%%%

\section{ACKNOWLEDGEMENTS}
\par The authors thank Xingen Liu, Minglang Hu, Wei Wu, and Yue-Wen Fang for stimulating discussions.
\par This work was supported by the Tokyo Tech World Research Hub Initiative (WRHI) Program of the Institute of Innovative Research, Tokyo Institute of Technology, the National Natural Science Foundation of China (Grants No.~51861145315, No.~11929401, No.~12074241, No.~52130204), the Science and Technology Commission of Shanghai Municipality (Grant No.~19010500500, No.~20501130600, No.~19DZ2270200), the Independent Research and Development Project of State Key Laboratory of Advanced Special Steel, Shanghai Key Laboratory of Advanced Ferro Metallurgy, Shanghai University (Grant No.~SKLASS 2020-Z07), Austrian Research Promotion Agency (FFG, Grant No.~870024, project acronym MagnifiSens), and High Performance Computing Center, Shanghai University.

\section{AUTHOR CONTRIBUTIONS}
W.R. and S.A.N conceived the study. X.M., S.A.N., and C.L. performed first-principles calculations of electronic structures, lattice dynamics, and ferroelectric properties. S.A.N. performed the analysis of structural stability and calculations of the Wannier functions. S.A.N. and X.M. investigated the properties of domain walls. S.A.N. analyzed the results and wrote the manuscript with the input from all authors. \\

\section{DATA AVAILABILITY}
The authors declare that all source data supporting the findings of this study are available within the article and the Supplementary information file.

\section{COMPETING INTERESTS}
The authors declare no competing interests.


\begin{thebibliography}{99}

\bibitem{ferro1}
Dawber, M., Rabe, K. M., Scott, J. F. Physics of thin-film ferroelectric oxides. \emph{Rev. Mod. Phys.} \textbf{77}, 1083--1130 (2005).

\bibitem{ferro2}
Scott, J. F. Applications of Modern Ferroelectrics. \emph{Science} \textbf{315}, 954--959 (2007).

\bibitem{thin1}
Bune, A. V. et al. Two-dimensional ferroelectric films. \emph{Nature} \textbf{391}, 874--877 (1998).

\bibitem{thin2}
Fong, D. D. et al. Ferroelectricity in ultrathin perovskite films. \emph{Science} \textbf{304}, 1650--1653 (2004).

\bibitem{thin3}
B\"oscke, T. S., M\"uller, J., Br\"auhaus, D., Schr\"oder, U., B\"ottger, U. Ferroelectricity in hafnium oxide thin films. \emph{Appl. Phys. Lett.} \textbf{99}, 102903 (2011).

\bibitem{thin4}
Martin, L. W. \& Rappe, A. M. Thin-film ferroelectric materials and their applications. \emph{Nat. Rev. Mater.} \textbf{2}, 16087 (2017).

\bibitem{thin5}
Wang, H. et al. Direct observation of room-temperature out-of-plane ferroelectricity and tunneling electroresistance at the two-dimensional limit. \emph{Nat. Commun.} \textbf{9}, 3319 (2018).

\bibitem{assembly1}
Geim, A. K. \& Grigorieva, I. V. Van der Waals heterostructures. \emph{Nature} \textbf{499}, 419--425 (2013).

\bibitem{assembly2}
Frisenda, R. et al. Recent progress in the assembly of nanodevices and van der Waals heterostructures by deterministic placement of 2D materials. \emph{Chem. Soc. Rev.} \textbf{47}, 53--68 (2018). 

\bibitem{boron1}
Stern, M. V. et al. Interfacial Ferroelectricity by van der Waals Sliding. \emph{Science} \textbf{372}, 1462--1466 (2021).

\bibitem{boron2}
Yasuda, K., Wang, X., Watanabe, K., Taniguchi, T., Jarillo-Herrero, P. Stacking-engineered ferroelectricity in bilayer boron nitride. \emph{Science} \textbf{372}, 1458--1462 (2021).

\bibitem{vdwreview}
Duong, D. L., Yun, S. J., Lee, Y. H. Van der Waals Layered Materials: Opportunities and Challenges. \emph{ACS Nano} \textbf{11}, 11803--11830 (2017).

\bibitem{thin33}
Shirodkar, S. N. \& Waghmare, U. V. Emergence of ferroelectricity at a metal-semiconductor transition in a 1T monolayer of MoS$_{2}$. \emph{Phys. Rev. Lett.} \textbf{112}, 157601 (2014).

\bibitem{thin44}
Fei, R. X., Kang, W., Yang, L. Ferroelectricity and phase transitions in monolayer group-IV monochalcogenides. \emph{Phys. Rev. Lett.} \textbf{117}, 097601 (2016).

\bibitem{wte2switch}
Fei, Z. et al. Ferroelectric switching of a two-dimensional metal. \emph{Nature} \textbf{560}, 336--339(2018).

\bibitem{polarmote2}
Yuan, S. et al. Room-temperature ferroelectricity in MoTe$_{2}$ down to the atomic monolayer limit. \emph{Nat. Commun.} \textbf{10}, 1775 (2019).

\bibitem{magneto1}
Ali, M. N. et al. Large, non-saturating magnetoresistance in WTe$_2$. \emph{Nature} \textbf{514}, 205--208 (2014).

\bibitem{supermote2}
Qi, Ya. et al. Superconductivity in Weyl semimetal candidate MoTe$_2$. \emph{Nat. Commun.} \textbf{7}, 11038 (2016)

\bibitem{zhijun}
Wang, Z.  et al. MoTe$_{2}$: A Type-II Weyl topological Metal. \emph{Phys. Rev. Lett.} \textbf{117}, 056805 (2016).

\bibitem{higherorder}
Wang, Z., Wieder, B. J., Li, J., Yan, B., Bernevig, B. A. Higher-Order Topology, Monopole Nodal Lines, and the Origin of Large Fermi Arcs in Transition Metal Dichalcogenides $X{\mathrm{Te}}_{2}$ ($X=\mathrm{Mo},\mathrm{W}$). \emph{Phys. Rev. Lett.} \textbf{123}, 186401 (2019).

\bibitem{polarmote}
Huang, F.-T. et al. Polar and phase domain walls with conducting interfacial states in a Weyl semimetal MoTe$_{2}$. \emph{Nat. Commun.} \textbf{10}, 4211 (2019).

\bibitem{wte1}
Sharma, P. et al. A room-temperature ferroelectric semimetal. \emph{Sci. Adv.} \textbf{5}, eaax5080 (2019).

\bibitem{cuinps}
Liu, F. et al. Room-temperature ferroelectricity in CuInP$_{2}$S$_{6}$ ultrathin flakes. \emph{Nat. Commun.} \textbf{7}, 12357 (2016).

\bibitem{inse1}
Zhou, Y. et al. Out-of-plane piezoelectricity and ferroelectricity in layered $\alpha$-In$_{2}$Se$_{3}$ nanoflakes. \emph{Nano Lett.} \textbf{17}, 5508--5513 (2017).

\bibitem{inse2}
Cui, C. et al. Intercorrelated In-Plane and Out-of-Plane Ferroelectricity in Ultrathin Two-Dimensional Layered Semiconductor In$_{2}$Se$_{3}$. \emph{Nano Lett.} \textbf{18}, 1253--1258 (2018).

\bibitem{renwei}
Liu, X., Pyatakov, A. P., Ren, W. Magnetoelectric Coupling in Multiferroic Bilayer VS$_{2}$. \emph{Phys. Rev. Lett.} \textbf{125}, 247601 (2020).

\bibitem{struc1}
Guthrie, D. H. \& Corbett, J. D. Synthesis and structure of an infinite-chain form of ZrI$_{2}$ ($\alpha$), \emph{J. Solid State Chem.} \textbf{37}, 256--263 (1981).

\bibitem{struc2}
Corbett, J. D. \& Guthrie, D. H. A second infinite-chain form of zirconium diiodide ($\beta$) and its coherent intergrowth with $\alpha$-zirconium diiodide. \emph{Inorg. Chem.}  \textbf{21}, 1747--1751 (1982).

\bibitem{struc3}
Guthrie, D. H. \& Corbett, J. D. Two zirconium iodide clusters. Hexazirconium dodecaiodide (Zr$_{6}$I$_{12}$) and cesium hexazirconium tetradecaiodide (CsZr$_{6}$I$_{14}$). \emph{Inorg. Chem.}  \textbf{21}, 3290--3295 (1982).

\bibitem{rabe}
Rabe K. M., Ahn, C. H., Triscone, J.‐M. Physics of Ferroelectrics: A Modern Perspective. Springer‐Verlag Berlin Heidelberg, Vol. 105 (2007).

\bibitem{liju}
Yang, Q., Wu, M., Li, J. Origin of Two-Dimensional Vertical Ferroelectricity in WTe$_{2}$ Bilayer and Multilayer. \emph{J. Phys. Chem. Lett.} \textbf{9}, 7160--7164 (2018).

\bibitem{ren1}
Liu, X. et al. Vertical ferroelectric switching by in-plane sliding of two-dimensional bilayer WTe$_{2}$. \emph{Nanoscale} \textbf{11}, 18575--18581 (2019).

\bibitem{prbzri2}
Zhang, T. et al. Ferroelastic-ferroelectric multiferroics in a bilayer lattice. \emph{Phys. Rev. B} \textbf{103}, 165420 (2021).

\bibitem{dwreview}
Catalan, G., Seidel, J., Ramesh, R., Scott, J. F. Domain wall nanoelectronics. \emph{Rev. Mod. Phys.} \textbf{84}, 119--156 (2012).

\bibitem{cdwreview}
Bednyakov, P. S., Sturman, B. I., Sluka, T. et al. Physics and applications of charged domain walls. \emph{npj Comput. Mater.} \textbf{4}, 65 (2018).

\bibitem{enhanced}
Sluka, T., Tagantsev, A. K., Damjanovic, D., Gureev, M., Setter, N. Enhanced electromechanical response of ferroelectrics due to charged domain walls. \emph{Nat. Commun.} \textbf{3}, 748 (2012).

\bibitem{dw1}
Meyer, B. \& Vanderbilt, D. Ab initio study of ferroelectric domain walls in PbTiO$_{3}$. \emph{Phys. Rev. B} \textbf{65}, 104111 (2002). 

\bibitem{dw2}
Ren, W. et al. Ferroelectric Domains in Multiferroic ${\mathrm{BiFeO}}_{3}$ Films under Epitaxial Strains. \emph{Phys. Rev. Lett.} \textbf{110}, 187601 (2013).

\bibitem{dw3}
Yang, Ya. et al. Improper ferroelectricity at antiferromagnetic domain walls of perovskite oxides. \emph{Phys. Rev. B} \textbf{96}, 104431 (2017).

\bibitem{quasigas}
Sluka, T., Tagantsev, A. K., Bednyakov, P., Setter, N. Free-electron gas at charged domain walls in insulating BaTiO$_{3}$. \emph{Nat. Commun.} \textbf{4}, 1808 (2013).

\bibitem{gas1}
Ohtomo, A. \& Hwang, H. A high-mobility electron gas at the LaAlO$_{3}$/SrTiO$_{3}$ heterointerface. \emph{Nature} \textbf{427}, 423--426 (2004). 

\bibitem{debyelength}
Sturman, B., Podivilov, E., Stepanov, M., Tagantsev, A., Setter, N. Quantum properties of charged ferroelectric domain walls. \emph{Phys. Rev. B} \textbf{92}, 214112  (2015).

\bibitem{junction}
Garcia, V. \& Bibes, M. Ferroelectric tunnel junctions for information storage and processing. \emph{Nat. Commun.} \textbf{5}, 4289 (2014).

\bibitem{expmote2}
Tamai, A. et al. Fermi Arcs and Their Topological Character in the Candidate Type-II Weyl Semimetal ${\mathrm{MoTe}}_{2}$. \emph{Phys. Rev. X} \textbf{6}, 031021 (2016).

\bibitem{expwte2}
Brown, B. E. The crystal structures of WTe$_{2}$ and high-temperature MoTe$_{2}$. \emph{Acta Cryst.} \textbf{20}, 268--274 (1966).

\bibitem{expwte22}
Mar, A., Jobic, S., Ibers, J. A. Metal-metal vs tellurium-tellurium bonding in WTe$_{2}$ and its ternary variants TaIrTe$_{4}$ and NbIrTe$_{4}$. \emph{J. Am. Chem. Soc.} \textbf{114}, 8963--8971 (1992).

\bibitem{bilbao1}
Aroyo, M. I., Perez-Mato, J. M., Orobengoa, D.,  Tasci, E., de la Flor, G., Kirov, A. Crystallography online: Bilbao Crystallographic Server. \emph{Bulg. Chem. Commun.} \textbf{43}, 183--197 (2011)

\bibitem{bilbao2}
Capillas, C. et al. SYMMODES: a software package for group-theoretical analysis of structural phase transitions. \emph{J.~Appl. Cryst.} \textbf{36}, 953--954 (2003).

\bibitem{vesta}
Momma, K. \& Izumi, F. VESTA3 for three‐dimensional visualization of crystal, volumetric and morphology data. \emph{J.~Appl. Crystallogr.} \textbf{44}, 1272--1276 (2011).

\bibitem{vasp}
Kresse, G. \& Hafner, J. Ab initio molecular dynamics for liquid metals. \emph{Phys. Rev. B} \textbf{47}, 558--561 (1993).

\bibitem{paw}
Blochl, P. E. Projector augmented-wave method. \emph{Phys. Rev. B} \textbf{50}, 17953 (1994).

\bibitem{qe}
Giannozzi, P. et al. QUANTUM ESPRESSO: a modular and open-source software project for quantum simulations of materials. \emph{J.~Phys.: Condens.Matter} \textbf{21}, 395502 (2009).

\bibitem{lda}
Ceperley, D. M. \& Alder, B. J. Ground State of the Electron Gas by a Stochastic Method. \emph{Phys. Rev. Lett.} \textbf{45}, 566--569 (1980).

\bibitem{pbe}
Perdew J. P., Burke, K., Ernzerhof, M. Generalized Gradient Approximation Made Simple. \emph{Phys. Rev. Lett.} \textbf{77}, 3865--3868 (1996).

\bibitem{pbesol}
Csonka, G. I. et al. Assessing the performance of recent density functionals for bulk solids. \emph{Phys. Rev. B} \textbf{79}, 155107 (2009).

\bibitem{corso}
Dal Corso, A. Pseudopotentials periodic table: From H to Pu. \emph{Computational Material Science} \textbf{95}, 337--350 (2014).

\bibitem{mp}
Monkhorst, H. J. \& Pack, J. D. Special points for Brillouin-zone integrations. \emph{Phys. Rev. B} \textbf{13}, 5188--5192 (1976).

\bibitem{grimme1}
Grimme, S. Semiempirical GGA-type density functional constructed with a long‐range dispersion correction. \emph{J. Comp. Chem.} \textbf{27}, 1787--1799 (2006).

\bibitem{grimme2}
Grimme, S., Antony, J., Ehrlich, S., Krieg, H. A consistent and accurate ab initio parametrization of density functional dispersion correction (DFT-D) for the 94 elements H-Pu. \emph{J. Chem. Phys.} \textbf{132}, 154104 (2010).

\bibitem{grimme3}
Grimme, S., Ehrlich, S., Goerigk, L. Effect of the damping function in dispersion corrected density functional theory. \emph{J. Comp. Chem.} \textbf{32}, 1456--1465 (2011).

\bibitem{dion1}
 Dion, M., Rydberg, H., Schr\"oder, E., Langreth, D. C., Lundqvist, B. I. Van der Waals Density Functional for General Geometries. \emph{Phys. Rev. Lett.} \textbf{92}, 246401 (2004).

\bibitem{dion2}
Rom\'an-P\'erez, G. \& Soler, J. M. Efficient Implementation of a van der Waals Density Functional: Application to Double-Wall Carbon Nanotubes. \emph{Phys. Rev. Lett.} \textbf{103}, 096102 (2009).

\bibitem{dion3}
Lee, K., Murray, \'E. D., Kong, L., Lundqvist, B. I., Langreth, D. C. Higher-accuracy van der Waals density functional. \emph{Phys. Rev. B} \textbf{82}, 081101(R) (2010).

\bibitem{optdion1}
Klime\ifmmode \check{s}\else \v{s}\fi{}, J., Bowler, D. R., Michaelides, A. Chemical accuracy for the van der Waals density functional. \emph{J. Phys.: Condens. Matter} \textbf{22}, 022201 (2009).

\bibitem{optdion2}
Klime\ifmmode \check{s}\else \v{s}\fi{}, J., Bowler, D. R., Michaelides, A. Van der Waals density functionals applied to solids. \emph{Phys. Rev. B} \textbf{83}, 195131 (2011).

\bibitem{ldau}
Liechtenstein, A. I., Anisimov, V. I., Zaanen, J. Density-functional theory and strong interactions: Orbital ordering in Mott-Hubbard insulators. \emph{Phys. Rev. B} \textbf{52}, R5467--R5470 (1995).

\bibitem{gw1}
Hedin, L. New Method for Calculating the One-Particle Green's Function with Application to the Electron-Gas Problem. \emph{Phys. Rev.} \textbf{139}, A796--A823 (1965).

\bibitem{gw2}
Shishkin, M. \& Kresse, G. Implementation and performance of the frequency-dependent $GW$ method within the PAW framework. \emph{Phys. Rev. B} \textbf{74}, 035101 (2006).

\bibitem{gw3}
Botti, S. \& Marques, M. A. L. Strong Renormalization of the Electronic Band Gap due to Lattice Polarization in the $GW$ Formalism. \emph{Phys. Rev. Lett.} \textbf{110}, 226404 (2013).

\bibitem{phonopy}
Togo, A. \& Tanaka, I. First principles phonon calculations in materials science. \emph{Scr. Mater.} \textbf{108}, 1--5 (2015).

\bibitem{dfpt}
Baroni, S., de Gironcoli, S., Dal Corso, A., Giannozzi, P. Phonons and related crystal properties from density-functional perturbation theory. \emph{Rev. Mod. Phys.} \textbf{73}, 515--562 (2001).

\bibitem{bornvasp}
Gajdo\ifmmode\check{s}\else \v{s}\fi{}, M., Hummer, K., Kresse, G., Furthm\"uller, J., Bechstedt, F. Linear optical properties in the projector-augmented wave methodology. \emph{Phys. Rev. B} \textbf{73}, 045112 (2006).

\bibitem{berry1}
King-Smith, R. D. \& Vanderbilt, D. Theory of polarization of crystalline solids. \emph{Phys. Rev. B} \textbf{47}, 1651--1654 (1993).

\bibitem{berry2}
Resta, R. Macroscopic polarization in crystalline dielectrics: The geometric phase approach. \emph{Rev. Mod. Phys.} \textbf{66}, 899--915 (1994).

\bibitem{wan90}
Mostofi, A. A. et al. An updated version of wannier90: A tool for obtaining maximally-localised Wannier functions. \emph{Comput. Phys. Commun.} \textbf{185}, 2309--2310 (2014).

\bibitem{nose}
Nos\'e, S. A unified formulation of the constant temperature molecular dynamics methods. \emph{J. Chem. Phys.} \textbf{81}, 511--519 (1984).

\bibitem{debye}
Debye P. \& H\"uckel E. Zur Theorie Der Elektrolyte. I. Gefrierpunktserniedrigung Und Verwandte Erscheinungen. \emph{Phys. Z.} \textbf{24},  185--206 (1923).  \\

\end{thebibliography}
\end{document}